%% file: main.tex
\documentclass{Interspeech}
\usepackage{color,soul}

\include{declarations}

\interspeechcameraready 

\title{\method: Phonetic-Acoustic Speech Tokenizer}

\author{Nadav}{Har-Tuv}
\author{Or}{Tal}
\author{Yossi}{Adi}

\affiliation{The School of Computer Science and Engineering}{}{}
\affiliation{The Hebrew University of Jerusalem, Israel}{}{}

\email{nadav.har-tuv1@mail.huji.ac.il}

\usepackage{comment}

\keywords{Speech Tokenization, Phonetic and Acoustic Tokens, Speech Language Models}

\begin{document}

\maketitle

\begin{abstract}
    We present \method, a novel end-to-end framework that jointly models phonetic information alongside signal reconstruction, eliminating the need for external pretrained models. 
    Unlike previous approaches that rely on pretrained self-supervised models, \method employs supervised phonetic data, directly integrating domain knowledge into the tokenization process via auxiliary tasks.
    Additionally, we introduce a streamable, causal variant of \method, enabling real-time speech applications. 
    Results demonstrate that \method surpasses existing evaluated baseline tokenizers across common evaluation metrics, including phonetic representation and speech reconstruction. 
    Notably, \method also achieves superior performance when serving as a speech representation for speech language models, further highlighting its effectiveness as a foundation for spoken language generation. 
    To foster further research, we release the full implementation.
    Code, model checkpoints, and samples see \url{pages.cs.huji.ac.il/adiyoss-lab/PAST}.
\end{abstract}

\input{intro}
\input{related}
\input{method}

\input{exp_setup}

\input{experiments}
\input{discussion}

\bibliographystyle{IEEEtran}
\input{main.bbl}

\end{document}

%% file: declarations.tex
\usepackage{etoolbox}
\usepackage{amsfonts}
\usepackage{amssymb}% http://ctan.org/pkg/amssymb
\usepackage{pifont}% http://ctan.org/pkg/pifont
\usepackage[T1]{fontenc} % add special characters (e.g., umlaute)
\usepackage[utf8]{inputenc} % set utf-8 as default input encoding
\usepackage{amsmath,cite,url}
\usepackage{graphicx}
\usepackage{color}
\usepackage{booktabs}
\usepackage{lineno}
\usepackage{xspace}
\usepackage{bbm}
\usepackage{amsopn}
\usepackage{cuted}
\usepackage{xspace}
\usepackage{hyperref}
\usepackage{xcolor}
\usepackage{cite}

\usepackage{bm}
\usepackage{multirow}

\newcommand{\vp}{\bm{p}}       \newcommand{\vph}{\hat{\bm{p}}}        
\newcommand{\vq}{\bm{q}}

\newcommand{\vx}{\bm{x}}       \newcommand{\vxh}{\hat{\bm{x}}}        
\newcommand{\vy}{\bm{y}}               
\newcommand{\vz}{\bm{z}}       \newcommand{\vzh}{\hat{\bm{z}}}        

\newcommand{\R}{\mathbb{R}}

\newcommand{\newpara}[1]{\vspace{0.02cm}\noindent \textbf{#1}}

\newtoggle{release}
\togglefalse{release}
\newif\ifLM \LMtrue \newcommand{\lm}{\ifLM language model (LM) \global\LMfalse\else LM \fi}
\newif\ifAR \ARtrue 
\newif\ifFM \FMtrue 
\newif\ifTTS \TTStrue

\newif\ifWER \WERtrue \newcommand{\wer}{\ifWER Word Error Rate (WER) \global\WERfalse\else WER \fi}
\newif\ifCER \CERtrue 
\newif\ifASR \ASRtrue 

\newif\ifRVQ \RVQtrue \newcommand{\rvq}{\ifRVQ Residual Vector Quantization (RVQ)\global\RVQfalse\else RVQ \fi}

\newif\ifFAD \FADtrue 
\newif\ifKLD \KLDtrue 
\newif\ifSLM \SLMtrue \newcommand{\slm}{\ifSLM Speech Language Models (SLM)\global\SLMfalse\else SLM \fi}

\newif\ifSSL \SSLtrue 
\newif\ifSSM \SSMtrue \newcommand{\ssm}{\ifSSM self-supervised models (SSL models)\global\SSMfalse\else SSL models \fi}
\newif\ifCTC \CTCtrue \newcommand{\ctc}{\ifCTC connectionist temporal classification (CTC) \global\CTCfalse\else CTC \fi}

\newif\ifPNMI \PNMItrue \newcommand{\pnmi}{\ifPNMI Phone-Normalized Mutual Information (PNMI)\global\PNMIfalse\else PNMI \fi}
\newcommand{\method}{\textsc{PAST}\xspace}

\usepackage{cite}
\usepackage{amsmath,amssymb,amsfonts}
\usepackage{algorithmic}
\usepackage{graphicx}
\usepackage{textcomp}
\usepackage{xcolor}

%% file: intro.tex
\section{Introduction}
\label{sec:intro}
Speech and audio language models have recently attracted significant attention in the research community, showcasing remarkable performance across a range of tasks~\cite{gslm, textless, audiolm, pgslm, slmfromrow, twist}. 
These models usually operate over acoustic tokens or phonetic speech tokens (also known as semantic tokens).
Acoustic tokenizers, such as EnCodec~\cite{encodec} and SoundStream~\cite{soundstream}, are designed for high-fidelity waveform reconstruction but are less ideal for language modeling without an external text supervision.
In contrast, phonetic tokenizers, such as those derived from quantizing wav2vec 2.0 ~\cite{wav2vec2} and HuBERT \cite{HuBERT} latent representations, primarily capture linguistic information~\cite{phoneme_analysing}. This makes them more suitable for sequential modeling.
However, they require an additional vocoder module ~\cite{hifigan} for speech synthesis, which increases complexity and can often degrade reconstruction quality.

Hybrid tokenizers, such as ~\cite{speechtokenizer, xcodec, moshi}, integrate phonetic and acoustic information into a unified representation.
This is achieved by extracting and distilling phonetic features from pretrained \ssm, e.g. WavLM~\cite{wavlm}.
Self-supervised representations exhibit correlation to phonetic content, e.g. measured by phoneme mutual information~\cite{HuBERT, phoneme_analysing}, but they are not explicitly optimized to capture it.
Despite showing promise, due to their dependence on pretrained SSL models, such hybrid approaches are potentially limited in their ability to fully capture the phonetic richness of the input.
That being said, explicit phonetic supervision could not only benefit the tokenizer's ability to capture phonetic content, it could also replace the dependence on pretrained \ssm and lower the computational cost.

\begin{figure}[t!]
    \centering
    \includegraphics[width=\linewidth]{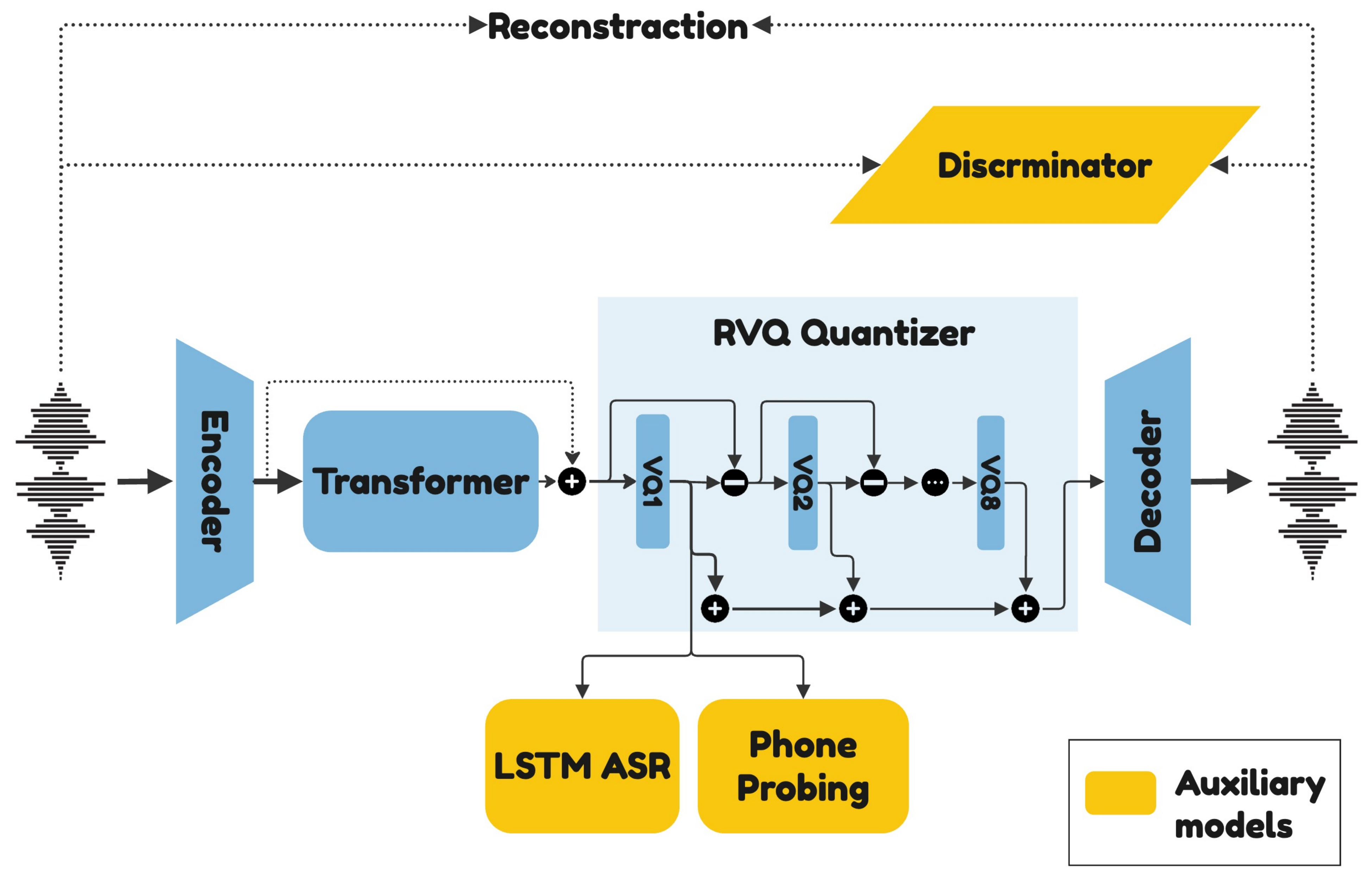}
    \caption{
        Schematic of the PAST pipeline. The auxiliary heads use the output of the first vector quantization module as input.
    }
    \label{fig:framwork}
    \vspace{-10pt}
\end{figure}

In this work, we introduce Phonetic-Acoustic Speech Tokenizer (\method), a novel end-to-end tokenizer that jointly captures both phonetic and acoustic representations without requiring external pretrained models or vocoders (see Figure~\ref{fig:framwork}).
In addition to the standard model, we present a variant of \method designed for streaming applications, which operates causally by relying only on past information.
\method achieves this by leveraging supervised auxiliary tasks, such as phoneme classification and automatic speech recognition, to incorporate phonetic information directly into the quantization process. 
This approach allows \method to outperform existing methods in phonetic and acoustic benchmarks while maintaining a simpler pipeline.

\noindent \textbf{Our contributions} are as follows:
\begin{enumerate}
    \item We propose a novel approach for jointly learning phonetic and acoustic representations using supervised data, eliminating the need for pretrained models or external vocoders.
    \item We achieve superior performance compared to hybrid tokenizers across both phonetic and acoustic benchmarks., demonstrating the effectiveness of our approach.
    \item We introduce a streaming-compatible variant of \method that operates causally, ensuring that it relies only on previous context, making it suitable for real-time speech applications.
    \item We open-source our implementation, including training and inference pipelines in addition to model checkpoints.
\end{enumerate}
\vspace{-5pt}

%% file: related.tex
\section{Related Work}
\label{sec:related_work}

% This work focuses on SLMs that operate over discrete latent representations.
Approaches to tokenization can be roughly categorized into phonetic tokenizers, acoustic tokenizers, and hybrid-tokenizers.

\noindent \textbf{Acoustic tokenizers} aim to compress speech into discrete representations optimized for high-fidelity reconstruction. 
EnCodec~\cite{encodec} and SoundStream~\cite{soundstream} have set the foundation for streamable high-fidelity audio compression by introducing \rvq.  
Subsequent works focused on enhancing fidelity through (i) improved quantization~\cite{hificodec,dac} (ii) improving latent representation structure~\cite{pyramidCodec} (iii) increasing training stability~\cite{audiodec} or (iv) utilizing spectral representations~\cite{funcodec, wavtokenizer}.

\noindent \textbf{Phonetic tokenizers}, on the other hand, quantize latent representations obtained by speech encoder model such as HuBERT~\cite{HuBERT}, wav2vec~2.0~\cite{wav2vec2}, or Whisper encoder~\cite{zeng2024scaling, radford2023robust} are trained to maximize mutual information across latent sequences of vectors, yet without explicit reconstruction constraints.
These types of models usually operate over continuous latent spaces, and hence require some discretization method like k-means, or learned clustering~\cite{last, messica2024nast}, to tokenize continuous embeddings into discrete tokens, enabling downstream modeling.
Many studies have demonstrated that these tokens effectively capture phonetic information\cite{gslm,pgslm,textless,twist}.
They exhibit strong correlations with phonemes~\cite{phoneme_analysing}, and their representations have been successfully used to improve performance in speech applications~\cite{tal2022systematic} and to train \slm, with a vocoder component employed to synthesize speech from generated tokens.

% \noindent \textbf{Distilled Tokenizers} leverage a pretrained phonetic teacher model as auxiliary objective for distillation guidance.
% These approaches aim to jointly capture phonetic information while maintaining signal reconstruction capabilities.
% \ort{We need to give a slightly more in-depth of related baselines here.}
% In SpeechTokenizer~\cite{speechtokenizer}, phonetic distillation is applied to the first \rvq codebook. 
% X-Codec~\cite{xcodec} incorporates phonetic features prior to the \rvq stage and employs a phonetic reconstruction loss post-\rvq to enhance representation quality. 
% In Mimi, Moshi's tokenizer~\cite{moshi} distills the phonetic representation directly into a single vector quantizer, which is then combined with acoustic tokens to achieve high-quality reconstruction.

\noindent \textbf{Hybrid tokenizers} aimed at merging phonetic and acoustic tokens. The common approach nowadays is to leverage a pretrained phonetic teacher model as auxiliary objective for distillation guidance.
These approaches aim to jointly capture phonetic information while maintaining signal reconstruction capabilities.
SpeechTokenizer~\cite{speechtokenizer} applies phonetic distillation to the first \rvq codebook, treating its indices as phonetic tokens while later codebooks encode acoustic details. 
This hierarchical approach enhances phonetic representation while maintaining signal quality.
X-Codec~\cite{xcodec} utilizes a pretrained SSL model during training and inference. 
It concatenate a low rank projection of the latent representation to the acoustic encoder output, in addition to applying an auxiliary distillation objective over the post-quantization latent. 
Mimi, the tokenizer used by Moshi~\cite{moshi}, is designed to condense phonetic information into a single quantized stream, which is then combined with the encoded acoustic quantized latent representation.

Despite their strong performance, approaches relying on pseudo-labels from \ssm have several limitations.
First, by relying on pseudo-supervision instead of supervised data, these approaches fail to fully utilize domain knowledge and may not align with explicit phonetic structures.
Second, these models require vast amounts of unlabeled speech data, making their training computationally expensive and less feasible for low-resource languages.
Finally, the inherent distillation process may encode cross-vector properties that are not directly relevant to phonetic representation, leading to redundancy and inefficiencies in the learned token space.

%% file: method.tex
\section{Method}
\label{sec:method}

\subsection{Problem Setup}
\label{sec:method:setup}
% Our model consists of 3 main building blocks: Encoder, Quantizer, and Decoder.
Our model is composed of three main components: Encoder, Quantizer, and Decoder. 
Given a waveform signal $\vx \in \mathbb{R}^{f_s \cdot t}$ of duration $t$[sec], sampled at $f_s$[Hz], the encoder transforms $\vx$ into a dense latent representation $\vz \in \mathbb{R}^{D \times T}$. Here, $T = f_r \cdot t$ denotes the temporal resolution of the latent space, determined by the frame rate $f_r$, and $D$ represents the latent dimension.
The Quantizer module then processes $\vz$, producing a quantized latent representation $\vzh \in \mathbb{R}^{D \times T}$. 
Finally, the Decoder reconstructs the original signal, yielding $\vxh \in \mathbb{R}^{f_s \cdot t}$.
To encourage capturing of phonetic content in the encoded latent representation we use paired phoneme and character-level transcription supervision with a set of auxiliary losses as described in subsection~\ref{sec:method:aux}.
Our goal is then to (i) minimize the reconstruction error between $\vx$ and $\vxh$, and (ii) to ensure that the encoded latent representation $\vz$ captures meaningful phonetic information.

\subsection{Model Architecture}
\label{sec:method:arch}
Our model is built on top of EnCodec~\cite{encodec} with an addition of a transformer encoder prior to the quantization module as depicted in Figure~\ref{fig:framwork}.
The Encoder block is comprised from a convolutional encoder module, same as EnCodec's encoder, followed by a transformer encoder module.
Experiments with and without the transformer encoder module are presented in Table~\ref{tab:ablation} and discussed in Section~\ref{sec:exp:abl_components}.

\noindent To enhance training stability, during training, the input to the quantization module is chosen from one of three modes of operations: (i) the output of the transformer block, with probability $p_{\text{trns.-only}}$; (ii) the output of the encoder (skip connection), with probability $p_{\text{skip-only}}$; or (iii) the average of (i) and (ii). During inference, the model is using the averaged representation only.
For the quantization module we employ Residual Vector Quantization.
The \rvq component contains $N_q$ sequential vector quantization (VQ) layers, which iteratively quantize $\vz$ along with its residuals.
Specifically, given $\vz\in \R^{D\times T}$, for every $\tau \in T$ the first VQ module replaces $\vz_{\tau}$ with the closest entry in a learned embedding table yielding $\vzh_1$.
% Then the process is repeated in the next VQ layers over the residue, i.e. for $i\in\{2,...,N_q\}: \text{VQ}_i\big(\vz - \sum_{j\in\{1,...,i-1\}}\vzh_j\big)=\vzh_i$.
Then the process is repeated in the next VQ layers over the residue, i.e. for $i\in\{2,...,N_q\}: \text{VQ}_i\big(\vz - \sum_{j\in[i-1]}\vzh_j\big)=\vzh_i$.
The \rvq module outputs $N_q$ quantized streams, which can be represented either as the quantized vectors $\vzh_i$ or as their corresponding indices $\vq_i \in \mathbb{N}^T$ in the embedding table of each $\text{VQ}_i$.
The Decoder component mirrors the convolutional encoder module, replacing the strided convolutions with transposed convolution layers.
The input to the decoder is $\vzh = \sum_{i\in [N_q]}\vzh_i$.

\noindent \textbf{Streamable configuration.} We introduce a streamable variant of our model with three key modifications: (i) causal convolutions utilizing left-only padding following~\cite{encodec}; 
% To enable real-time processing, we introduce a streamable variant of our model with three key modifications: (i) causal convolutions, where all padding is applied before the first time step, following EnCodec~\cite{encodec}; 
% (ii) unidirectional LSTM, replacing the bidirectional LSTM to enforce left-to-right processing; 
(ii) unidirectional LSTM and (iii) causal attention.
% and (iii) causal attention, restricting the transformer’s receptive field to past and present frames only.
% This ensures that the model operates with low latency, requiring only a $20$ms look-ahead window, over the audio signal.
This setup requires a $20$ms look-ahead window over the audio signal.

\subsection{Auxiliary Heads}
\label{sec:method:aux}
To encourage the encoding of phonetic information, we incorporate auxiliary heads and training objectives that operate over the first quantized output stream $\vzh_1$. 
This approach aims to replace the distillation of pseudo-phonetic tokens with direct supervision using target character transcriptions and phonemes.

\noindent \textbf{CTC character match.}
\label{sec:method:aux:ctc}
\noindent The CTC auxiliary head takes $\vzh_1\in\R^{D\times T}$ as input and outputs a distribution over the set of all characters $M$ at each entry $\vy \in \R^{|M| \times T}$.
The module is constructed from a linear projection from $D$ to a hidden dimension $h$ followed by a single-layer BiLSTM, and a linear projection from $h$ to $|M|$. 
A \ctc~\cite{ctc} loss, denoted as $\mathcal{L}_\text{ctc}=\text{CTC}(\vy| \text{chars})$, is then applied to align the predicted sequence with the transcription target.

\noindent \textbf{Phoneme classification.}
\label{sec:method:aux:phone}
\noindent The second auxiliary head is a simple linear projection that takes $\vzh_1$ as input and outputs a distribution over the set of all phonemes $P$ at each entry $\vph \in \R^{|P| \times T}$.
The auxiliary head is trained using a cross-entropy objective, denoted as $\mathcal{L}_{\text{phn}} = \text{CE}(\vph, \vp)$. 

\subsection{Training Objective}
\label{sec:method:loss}
We use the reconstruction training objective as defined in EnCodec's~\cite{encodec} paper, with the addition of two auxiliary terms.
EnCodec contains multiple training objectives, working in a weighted combination to optimize signal reconstruction. 
As we do not change the recommended weighted configuration of the involved objectives, we will refer to the overall EnCodec objective term as $\mathcal{L}_\text{EnCodec}$ and refrain from defining all of the components and notations here.
For further details, please refer to~\cite{encodec}.
The overall training objective minimizes the following:
\begin{equation}
    \mathcal{L} = \lambda_\text{ctc}\mathcal{L}_\text{ctc} + \lambda_\text{phn}\mathcal{L}_\text{phn} + \mathcal{L}_\text{EnCodec}
\end{equation}
where $\lambda_\text{ctc}$ and $\lambda_\text{phn}$ control the weights of the CTC loss $\mathcal{L}_\text{ctc}$ and phoneme loss $\mathcal{L}_\text{phn}$, respectively.

%% file: exp_setup.tex
\begin{table}[t!]
\caption{Comparison based on phonetic information.}
\centering
\renewcommand{\arraystretch}{1.2}
\resizebox{0.95\columnwidth}{!}{
\begin{tabular}{@{}l|c|cc|cc@{}}
\toprule
\textbf{Tokenizer}       & \textbf{PNMI $\uparrow$}  & \multicolumn{2}{c|}{\textbf{ABX $\downarrow$}} & \multicolumn{2}{c}{\textbf{WER $\downarrow$}} \\ 
                         &                           & \textbf{Within}       & \textbf{Across}          & \textbf{Clean}   & \textbf{Other} \\ \midrule
D. HuBERT 500            & 0.67                      & 3.91                  & 4.73                     & 11.3          & 24.7              \\ \midrule
SpeechTokenizer          & 0.72                      & 3.43                  & 4.50                     & 18.5          & 41.3              \\
X-Codec                  & 0.40                      & 9.42 & 12.6 & 17.1          & 37.1              \\ 
\textbf{PAST}            & \textbf{0.75}             & \textbf{2.82}         & \textbf{3.54}            & 15.7          & 36.8              \\ 
\textbf{PAST - Streamable}  & 0.74                      & 3.05                  & 3.89                     & \textbf{14.3} &  \textbf{32.3}    \\ \bottomrule
\end{tabular}}
\vspace{-0.35cm}
\label{tab:phonetic}
\end{table}

\section{Experimental Setup}
\label{sec:exp_setup}

\subsection{Data}
\label{sec:exp_setup:data}
We use all training subsets of LibriSpeech~\cite{Librispeech} and TIMIT~ \cite{timit} for our training set, yielding a total of $965$ hours of raw audio.
To improve training efficiency and avoid redundant padding, we sample $3$-second audio segments from each data sample. 
This segmentation is done purely for efficiency and is not an inherent part of our approach.
To achieve this, we obtain character-level alignment for paired text transcriptions using pretrained Wav2Vec2~\cite{wav2vec2} model.
For each sampled audio segment, Wav2Vec2 outputs a character distribution per temporal entry, from which we select the most probable alignment based on log-likelihood.
For the phoneme classification, we utilize the phonetic transcriptions provided by the TIMIT dataset. We sample instances in a 9:1 ratio (LS:T), resulting in $10\%$ of batch samples having phoneme supervision.

\subsection{Model Configuration}
\label{sec:exp_setup:model}

The encoder and decoder architectures follow the configuration from EnCodec~\cite{encodec} considering temporal downscaling with ratios $[8, 5, 4, 2]$, resulting in a frame rate of $50$ Hz. The latent dimensionality $D$ is set to $128$. The transformer module includes $8$ layers with a hidden size of $768$, $16$ attention heads, and a feed-forward size of $2048$. 
During training, the sampling probabilities for the transformer skip-connection  are $p_{\text{trns.-only}} = 0.3$ and $p_{\text{skip-only}} = 0.1$. 
The transformer processes up to $150$ frames ($3$ seconds), and longer sequences are divided into chunks with $1$-second overlap, averaging the results over the overlapping regions. The auxiliary loss weights are $\lambda_\text{ctc} = 12$ and $\lambda_\text{phn} = 5$, while the reconstruction loss weights follow the configuration defined in~\cite{encodec}. 
The \rvq contains $8$ codebooks, each with $1024$ entries. The dimension of the CTC auxiliary head is $h=512$.
The model has $185$M parameters, with $68$M allocated to the transformer and $4$M to the auxiliary heads.

\subsection{Training Configuration}
\label{sec:exp_setup:training}
Training is performed using two \textit{NVIDIA A100} GPUs with a batch size of $80$ for a total of $400,000$ steps, using the ADAM optimizer with $[0.5, 0.9]$ betas and no weight decay. 
The learning rate was managed using a cosine decay scheduler, starting at $3\cdot 10^{-4}$ and decreasing gradually to zero with a warm-up phase of $4,000$ steps.

\subsection{Evaluation Metrics}
\label{sec:exp_setup:eval}

\subsubsection{Signal Reconstruction Quality Metrics}
\label{sec:exp_setup:eval:recon}

The following are computed on the LibriSpeech clean-test set.

\noindent \textbf{Virtual Speech Quality Objective Listener} (ViSQOL)~\cite{chinen2020visqol} evaluates the reconstruction quality by comparing the spectral and temporal features of the reconstructed signal to the source signal. It produces an approximation of Mean Opinion Score.

\noindent \textbf{Scale-Invariant Signal-to-Noise Ratio} (SISNR) measures the similarity between the original and reconstructed signals by quantifying the ratio of target signal energy to residual noise.

\noindent \textbf{PESQ} (Perceptual Evaluation of Speech Quality) \cite{pesq} assesses the perceptual degradation of reconstructed signals, following the ITU-T P.862.2 wideband recommendation.

\subsubsection{Phonetic Information Evaluation}
\label{sec:exp_setup:eval:phonetic}

\noindent \textbf{\pnmi} \cite{HuBERT} quantifies the percentage of uncertainty about a given phone label $Y$ eliminated after observing a token $X$. It measures the mutual information  $I(X;Y)$, normalized by the entropy of $Y$. Higher values correspond to better phonetic encoding.
In our evaluation, \pnmi is computed on the tokenizer’s first \rvq codebook token $\vq_1$ using the TIMIT test-set.

\noindent \textbf{ABX} metric \cite{abx} measures the model’s ability to discriminate phonetic contrasts by comparing triplets of sounds $A$, $B$, and $X$. 
Lower ABX error rates reflect better phonetic preservation. 
We compute ABX on the reconstructed representation after the RVQ, evaluating both within-speaker and across-speaker contexts. 
This evaluates how well the tokenizer retains phonetic distinctions through the quantization process.

\noindent \textbf{\wer} measures the accuracy of generated transcriptions with respect to a reference transcription, where lower values indicate better performance.
\wer is computed using the DASB benchmark~\cite{dasb} on the discrete tokens from all codebooks $\vq$. 
Training and validation were performed on LibriSpeech \textit{train-clean-100} and \textit{dev-clean} subsets, while testing was conducted on the \textit{test-clean} and \textit{test-other} subsets.

\begin{table}[t!]
\caption{Signal reconstruction evaluation.}
\centering
\resizebox{0.85\columnwidth}{!}{
\renewcommand{\arraystretch}{1} % Adjust row height
\setlength{\tabcolsep}{7pt} % Adjust column width
\begin{tabular}{@{}l|ccc@{}}
\toprule
\textbf{Tokenizer}       & \textbf{SISNR $\uparrow$} & \textbf{VISQOL $\uparrow$} & \textbf{PESQ $\uparrow$} \\ \midrule
EnCodec                  & 7.49             & 4.48            &    3.88     \\ \midrule
SpeechTokenizer          & 0.44             & 4.38            &    3.15   \\
X-Codec                  & -7.12              & \textbf{4.46}   &    3.33   \\ 
\textbf{PAST}            & \textbf{4.84}    & 4.40            &    \textbf{3.55}   \\ 
\textbf{PAST - Streamable} & 3.90             & 4.37     &    3.40      \\ \bottomrule
\end{tabular}}
\label{tab:acoustic}
\vspace{-0.35cm}
\end{table}

\subsubsection{Speech Language Modeling Evaluation}
\label{sec:exp_setup:eval:slm}
In order to compare the observed methods, we train an identical backbone \lm for each of the observed tokenizers.
We leverage the base architecture of the AudioGen model \cite{audioGen} using the delay pattern, as suggested in MusicGen~\cite{musicGen}. 
We use a $300$M parameter model configuration and train on $10$-second audio segments for $300$k update steps with batch size of $256$, using all train subsets of LibriSpeech without any additional conditioning, text or prompts.
Equipped with this LM, we measure the \textbf{SWUGGY} metric~\cite{swuggy}. This metric evaluates the model’s ability to assign higher likelihood to valid words over pseudo-words by testing both Inter (within-vocabulary) and OOV (out-of-vocabulary) categories.
The probabilities are derived solely from the distribution of the first codebook, as it encapsulates the majority of the phonetic information in the output.

%% file: experiments.tex
\section{Results}
\label{sec:exp}
\vspace{-0.1cm}
\subsection{Baseline Comparison}
\label{sec:exp:baseline_compare}
\vspace{-0.1cm}
We compare \method with two baseline hybrid models, SpeechTokenizer and X-Codec, on both reconstruction and phonetic information metrics.
For context, we use EnCodec as an acoustic topline, since its objective is purely signal fidelity without phonetic supervision.
We have discarded Mimi~\cite{moshi} from this comparison as our reproduced results, using the published model, were far from being on par with the observed baselines.
We also compare \method with two top-line models: k-means over HuBERT for phonetic representation, and EnCodec for acoustic reconstruction. 
First, we evaluate the observed methods over a set of phonetic metrics.
Results depicted in Table~\ref{tab:phonetic} suggests that \method outperforms the observed baselines across all observed metrics, even surpassing the topline on several key aspects.
These results demonstrate that direct supervision effectively captures phonetic information, eliminating the need for distillation from pretrained SSL models. 

Next, we evaluate \method's signal generation capabilities. Table~\ref{tab:acoustic} presents the results for reconstruction quality evaluation. \method significantly surpasses the observed baselines, further showcasing its ability to balance phonetic richness with acoustic quality. We speculate that X-Codec's notably low SISNR is due to the absence of a point-wise metric loss objective over the reconstructed waveform, leading to misalignment in signal reconstruction. Moreover, Tables~\ref{tab:phonetic},~\ref{tab:acoustic} also highlight \method's superiority while utilizing a causal modeling configuration which allows for streaming capabilities.

\begin{table}[t!]
\caption{Evaluation of SLM performance across tokenizers.}
\centering
\resizebox{0.50\columnwidth}{!}{
\begin{tabular}{@{}l|cc@{}}
\toprule
\textbf{Tokenizer}  & \multicolumn{2}{c}{\textbf{sWUGGY $\uparrow$}}   \\ 
                         &  \textbf{Inter}  & \textbf{Oov}              \\ \midrule
EnCodec                  & 56.3             & 53.7                      \\
D. HuBERT 500            & 67.9             & 55.4                      \\  \midrule
SpeechTokenizer          & 63.7             & 55.6                      \\
X-Codec                  & 55.1             & 52.9                      \\ 
\textbf{PAST}            & \textbf{71.8}    & \textbf{57.5}             \\
\textbf{PAST - Streamable} & 70.2           & 56.3                      \\ \bottomrule
\end{tabular}}
\label{tab:slms}
\vspace{-0.35cm}
\end{table}

\subsection{Speech Language Modeling}
\label{sec:exp:slms}
The following experiment revolves around the evaluation of the observed methods as speech tokenizers, where we train an identical \lm as specified in subsection~\ref{sec:exp_setup:eval:slm} using each method to tokenize the audio signals to discrete token sequences.
Results summarized in Table~\ref{tab:slms} shows that \method notably surpasses all other observed models, further emphasizing the advantages of our data-driven approach. 
% We also measured sBLIMP \cite{dunbar2021zero}, a metric evaluating syntactic and phonetic consistency by distinguishing between grammatically correct and incorrect sentences. All observed methods performed comparable or slightly better than the random $50.0$ score equivalent.

\subsection{Component Analysis}
\label{sec:exp:abl_components}

Table~\ref{tab:ablation} highlights the impact of the different components that compose \method and reveals two main trends.  
First, including the suggested auxiliary heads and their corresponding training objectives is crucial for encapsulating phonetic information in the learned latent space, with CTC being the most crucial of the two, as reflected in the ABX scores.  
Since the CTC objective optimizes all possible alignments between the latent representation and the character transcription, we hypothesize that it encourages the latent space to encode phonetic information in a structured manner across time steps.  
Second, considering both auxiliary objectives without the transformer module leads to a significant difference across all metrics. 
However, including it improves sequence modeling and signal reconstruction, as reflected in the ABX and SISNR trends.  

Practically, adding the transformer module and naively cascading it after the encoder leads to vanishing gradients, causing divergence during training.  
Table~\ref{tab:skip_abl} highlights the importance of including the suggested skip connection dropout during training.  
By design, we utilize small auxiliary heads with limited sequence modeling capacity to ensure that certain modeling capabilities are captured at the latent representation level.  
Our motivation for adding the transformer component is to enhance the model's sequence modeling capacity.  
Without dropout on the skip connections, the model effectively bypassed the transformer encoder, leading to similar performance as the configuration where the transformer is entirely omitted, as shown in Table~\ref{tab:ablation}.  
Hence, imposing the dropout constraint empirically shows notable improvement and prevents this phenomenon.

\begin{table}[t!]
\caption{Ablation study over model components.}
\centering
\renewcommand{\arraystretch}{1} % Adjust row height
\resizebox{1.0\columnwidth}{!}{
\begin{tabular}{@{}ccc||c|cc|cc||c@{}}
\toprule
\textbf{Trns}   & \textbf{Phn} & \textbf{CTC} & \textbf{PNMI $\uparrow$} & \multicolumn{2}{c|}{\textbf{ABX $\downarrow$}} & \multicolumn{2}{c||}{\textbf{WER $\downarrow$}} & \textbf{SISNR $\uparrow$}       \\ 
        &  \textbf{clf}   &     &                          & \textbf{W.in}          & \textbf{Ac.}     & \textbf{Cl.}    & \textbf{Oth.}       &             \\ \midrule
    X   &   X    &  X  &        0.32          & 21.2           &      26.7       &    87.3          &  93.7            &   7.49  \\
    V   &   X    &  X  &        0.35          & 19.4           &      25.5       &    93.8          &  96.6            &   7.02  \\
    V   &   V    &  X  &        0.71          & 16.0           &      22.3       &    43.0          &   71.8           &   5.74  \\ 
    V   &   X    &  V  &        0.50          & 4.42           &      5.51       &    15.2          &  35.8            &   4.94  \\
    X   &   V    &  V  &        0.73          & 3.48           &      4.28       &    15.8          &  36.6            &   5.24  \\ \midrule
    V   &   V    &  V  &        0.75          & 2.82           &      3.54       &    15.7          &  36.8            &   4.84  \\ \bottomrule
\end{tabular}}
\label{tab:ablation}
\end{table}

\begin{table}[t!]
\caption{Impact of transformer skip-connection dropout.}
\centering
\renewcommand{\arraystretch}{1} % Adjust row height
\resizebox{0.96\columnwidth}{!}{
\begin{tabular}{@{}l||c|cc|cc||c@{}}
\toprule
\textbf{Mode} & \textbf{PNMI $\uparrow$} & \multicolumn{2}{c|}{\textbf{ABX $\downarrow$}} & \multicolumn{2}{c||}{\textbf{WER $\downarrow$}} & \textbf{SISNR $\uparrow$}       \\ 
                    &               & \textbf{W.in} & \textbf{Ac.}      & \textbf{Cl.}    & \textbf{Oth.}   &             \\ \midrule
Skip - No Drop      &    0.74       & 3.93          & 4.96              & 15.5            & 36.5            & 5.24  \\
\textbf{PAST}       &    0.75       & 2.82          & 3.54              & 15.7            & 36.8            &  4.84       \\ \bottomrule
\end{tabular}}
\label{tab:skip_abl}
\vspace{-0.35cm}
\end{table}

%% file: discussion.tex
\section{Discussion}
\label{sec:discussion}
We introduce \method, a unified phonetic-acoustic novel speech tokenizer that integrates supervised phonetic information into the tokenization process while maintaining high-fidelity reconstruction.
Unlike existing approaches that rely on SSL models and external vocoders, \method directly incorporates phonetic supervision, producing a representation that is both semantically meaningful and acoustically precise.
Our results demonstrate that \method surpasses SOTA tokenizers in phonetic representation, speech reconstruction, and speech language modeling. 
Notably, its supervised design eliminates the need for pretrained SSL models.
Furthermore, the proposed streamable variant extends \method’s applicability to real-time speech applications.
Although \method offers several benefits, its reliance on labeled phonetic data limits its scalability to a multilingual setting. 
Future work will focus on adapting \method to a multilingual setting.

\newpara{Acknowledgments.} This research work was supported by ISF grant 2049/22.